\documentclass[12pt,english]{article}
\pdfoutput=1
\usepackage[T1]{fontenc}
\usepackage{amssymb}
\usepackage{amsmath}
\usepackage{cite}
\usepackage{epsfig}
\usepackage[unicode=true,
 bookmarks=true,bookmarksnumbered=false,bookmarksopen=false,
 breaklinks=false,pdfborder={0 0 1},backref=false,colorlinks=true]
 {hyperref}
\hypersetup{pdftitle={Entanglement and the Double Copy},
 pdfauthor={Clifford Cheung, Grant N. Remmen},
 citecolor=black,linkcolor=black,urlcolor=black}
\usepackage{breakurl}
\usepackage[hang,flushmargin]{footmisc}
\usepackage{breakurl}
\usepackage{color}

\makeatletter
\def\simgt{\mathrel{\lower2.5pt\vbox{\lineskip=0pt\baselineskip=0pt
           \hbox{$>$}\hbox{$\sim$}}}}
\def\simlt{\mathrel{\lower2.5pt\vbox{\lineskip=0pt\baselineskip=0pt
           \hbox{$<$}\hbox{$\sim$}}}}
\makeatother

\usepackage{mathtools}

\newcommand{\verteq}{\rotatebox{90}{$=$}}

\newcommand{\be}{\begin{equation}}
\newcommand{\ee}{\end{equation}}
\newcommand{\bea}{\begin{eqnarray}}
\newcommand{\eea}{\end{eqnarray}}
\newcommand{\eq}[2]{\be\begin{aligned}#1 \label{#2}\end{aligned}\ee}

\newcommand{\Ref}[1]{Ref.~\cite{#1}}

\newcommand{\Eq}[1]{Eq.~\eqref{#1}}

\newcommand{\Sec}[1]{Sec.~\ref{#1}}

\newcommand{\bra}[1]{\langle #1 |}
\newcommand{\ket}[1]{| #1 \rangle}

\newcommand{\Hil}{{\cal H}}

\begin{document}

\interfootnotelinepenalty=10000
\baselineskip=18pt

\hfill CALT-TH-2020-003
\hfill

\vspace{2cm}
\thispagestyle{empty}
\begin{center}
{\LARGE \bf
Entanglement and the Double Copy
}\\
\bigskip\vspace{1cm}{
{\large Clifford Cheung${}^{a}$ and Grant N. Remmen${}^{b}$}
} \\[7mm]
 {\it ${}^a$Walter Burke Institute for Theoretical Physics\\[-1mm]
    California Institute of Technology, Pasadena, CA 91125 \\[1.5 mm]
 ${}^b$Center for Theoretical Physics and Department of Physics \\[-1mm]
     University of California, Berkeley, CA 94720 and \\[-1mm]
     Lawrence Berkeley National Laboratory, Berkeley, CA 94720} \let\thefootnote\relax\footnote{\noindent e-mail: \url{clifford.cheung@caltech.edu}, \url{grant.remmen@berkeley.edu}} \\
 \end{center}
\bigskip
\centerline{\large\bf Abstract}
\begin{quote} \small
We construct entangled states of gluons that scatter exactly as if they were gravitons.  Operationally, these objects implement the double copy at the level of the wave function.  Our analysis begins with a general ansatz for a wave function characterizing gluons in two copies of ${\rm SU}(N)$ gauge theory.  Given relatively minimal assumptions following from permutation invariance and dimensional analysis, the three- and four-particle wave functions generate scattering amplitudes that automatically coincide exactly with gravity, modulo normalization.  For five-particle scattering the match is not automatic but imposing certain known selection rules on the amplitude is sufficient to uniquely reproduce gravity. The resulting amplitudes exhibit a color-dressed and permutation-invariant form of the usual double copy relations.
We compute the entanglement entropy between the two gauge theory copies and learn that these states are maximally-entangled at large $N$.   Moreover, this approach extends immediately to effective field theories, where Born-Infeld photons and Galileons can be similarly recast as entangled gluons and pions.  
\end{quote}

\setcounter{footnote}{0}

\newpage
\tableofcontents
\newpage

\section{Introduction}

The study of scattering amplitudes has unveiled a striking duality linking gravity and gauge theory at weak coupling.  Summarized by the schematic equation,
\eq{
{\rm graviton} \quad \leftrightarrow \quad  {\rm gluon} \otimes {\rm gluon},
}{eq:double_copy_eq}
the so-called ``double copy'' is a mechanical prescription for building tree and loop graviton amplitudes from products of their gauge theory cousins.  Signs of this squaring relation first arose in string theory under the guise of the KLT formula \cite{KLT}, which equates certain products of tree-level Yang-Mills (YM) amplitudes to those of gravitons, dilatons, and two-form fields. Decades later, the double copy was understood as but one facet of a mysterious duality between color and kinematics \cite{BCJ,Bern:2010ue,Bern:2019prr}, ultimately leading to the breakthrough application of these ideas to cutting-edge multiloop gravitational amplitudes relevant to everything from supergravity \cite{Bern:2012uf,Bern:2018jmv} to the black hole inspiral problem \cite{Bern:2019nnu,Bern:2019crd}.

While \Eq{eq:double_copy_eq} holds at the level of scattering amplitudes, a natural question is whether it signifies some deeper connection at the level of states in Hilbert space. Indeed, if there exists an explicit mapping between graviton and gluon states, deciphering it would immediately provide a concrete weak-weak dual description of gravity purely in terms of gauge theory. In this paper, we initiate an investigation of this possibility.  Crucially, decoding a putative state-to-state map directly from the KLT or BCJ double copy is no simple task.  The KLT relations apply to color-ordered amplitudes, which conveniently form a basis for all amplitudes but unfortunately do not themselves correspond to the scattering of physical, color-dressed states. On the other hand, the BCJ double copy is implemented at the level of auxiliary kinematic numerators rather than on-shell amplitudes.

For all of these reasons, we pursue a more bottom-up approach.
To begin, we consider a general superposition of entangled gluons from two decoupled copies of ${\rm SU}(N)$ gauge theory in general spacetime dimension, hereafter denoted as $ {\rm YM} \otimes \overline{\rm YM}$.   
Our aim is to construct an entangled superposition of $n$ gluons from each copy whose associated scattering amplitude is {\it identical} to that of $n$ gravitons.
This ``entanglement ansatz'' is quite general but still conforms to three basic assumptions: {\it i}) the coefficients of the superposition are polynomial functions of the external momenta, {\it ii}) gluon color indices are contracted across but not within each copy, and {\it iii}) the resulting object is permutation invariant on the external kinematics.  Here, {\it i}) and {\it ii}) are assumed to reduce the complexity of our analysis but---as we will discuss later---also to mirror the mathematical form of the KLT double copy and Kaluza-Klein reduction.  Condition {\it iii}) is required so that the entangled state exhibits the underlying Bose symmetry required of gravitons.  Note that we work in all-in formalism throughout, so this entangled state is peculiar in that it contains both positive- and negative-energy modes.

Next, we ask, under what conditions does this entangled gluon state, evolving by the independent dynamics prescribed by each gauge theory copy, scatter exactly as if it were a collection of gravitons?  As it turns out, for three- and four-particle scattering, {\it any} choice of entanglement ansatz produces the known amplitudes for gravitons, dilatons, and two-forms up to overall normalization.  For the three-particle case this is of course completely expected given that the amplitude is uniquely fixed by Poincar\'e invariance \cite{Cachazo+Benincasa,Elvang:2015rqa,Cheung:2017pzi}.  However, what is more surprising is that the same also holds for four-particle scattering, where a match to gravity requires highly nontrivial cancellations among the many spurious poles that contaminate products of gluon amplitudes.  
For five-particle scattering, the entanglement ansatz does not automatically reproduce the known gravity amplitudes and the coefficients in the superposition must be chosen accordingly for a match.  At the same time, we observe that this choice of coefficients is specified automatically if we impose selection rules on the scattering amplitude as mandated by dilaton parity and ${\rm U}(1)$ duality \cite{Bern:2019prr}.  

Because these entangled gluon wave functions are manifestly permutation invariant, their corresponding amplitudes are as well.  Thus, our final expressions for these gravitational amplitudes exhibit an elegant form of the KLT relations that is manifestly permutation invariant and written in terms of color-dressed amplitudes.   We also present a set of remarkably simple entangled gluon states that reproduce three-, four-, and five-graviton scattering.  

Having recast gravitons as certain entangled states of gluons, it is natural to compute the entanglement entropy associated with this wave function.  By tracing over one gauge theory copy, we learn that these entangled gluon states are maximally entangled at large $N$.

We present all of our results in terms of the standard double copy relation between gauge theory and gravity.  However, encoding of the double copy in terms of entangled states can also be applied to certain effective field theories, e.g.,
\eq{
{\rm Galileon} &\quad \leftrightarrow \quad {\rm pion}  \otimes {\rm pion}  \\ 
\textrm{Born-Infeld} &\quad \leftrightarrow \quad {\rm gluon} \otimes {\rm pion},
}{}
which satisfy precisely the same KLT and BCJ relations discovered in gauge theory and gravity~\cite{Chen:2013fya,Carrasco:2016ldy,flavor-kinematics} and which are in fact inextricably interwoven with these theories~\cite{Cachazo:2014xea,Cheung:2017ems}.

This paper is organized as follows.
In \Sec{sec:ansatz} we construct the general entanglement ansatz and deduce a set of simple relations that enforce manifest permutation invariance on the wave function.
We then compute the scattering amplitudes associated with these entangled states in \Sec{sec:amplitudes}, using dimensional analysis to fix the mass dimension of the wave function coefficients.  
In Secs.~\ref{sec:3pt}, \ref{sec:4pt}, and \ref{sec:5pt}, we explicitly construct the permutation-invariant wave functions of entangled gluons for three-, four-, and five-particle states, respectively, and compare directly to gravity. 
We consider the entanglement structure of these states in \Sec{sec:entanglement} and conclude in \Sec{sec:conclusions}.

\section{Entanglement Ansatz}\label{sec:ansatz}

A multiparticle state of $n$ gravitons will be denoted by 
\eq{
 |p_1\rangle   | p_2 \rangle  \cdots  | p_n\rangle = |p\rangle \in \Hil_{\rm gravity} ,
}{eq:grav_ket}
where $p$ is an $n$-component vector whose elements are the external spacetime momentum vectors $p_i$ for $i=1,2,\ldots, n$.  Throughout this paper, any discussion of the ``gravity theory'' or ``gravitons'' will, unless stated otherwise, refer implicitly to the extended theory of the graviton, dilaton, and two-form field that appears in the double copy \cite{Bern:2019prr} and that comprises the field theory zero modes of the closed string.  

We also define a multiparticle state of $n$ gluons in YM theory by
\eq{
|p_1, a_1\rangle   | p_2,a_2 \rangle  \cdots  | p_n, a_n\rangle = |p,a\rangle  \in \Hil_{\rm YM} ,
}{eq:gauge_ket}
where $a$ is an $n$-component vector whose elements are the color indices $a_i$ for $i=1,2,\ldots, n$.    Note that all of these states implicitly carry external polarization labels, which we suppress throughout.\footnote{Amusingly, at fixed momenta the graviton state $\ket{p}$ in four spacetime dimensions is literally a string of $n$ qubits indexed by the external helicities.  This same is also true for gluons, albeit with additional color structure. }  We emphasize that the relative ordering of $p$ and $a$ in \Eq{eq:gauge_ket} is critically important since it dictates which color indices are assigned to which external leg.

By Bose symmetry, the multigraviton and multigluon states are permutation invariant, so
\eq{
|p\rangle = |\sigma (p)\rangle \qquad {\rm and} \qquad  |p ,a\rangle = |\sigma (p),\sigma (a)\rangle,
}{eq:Bose}
where the permutation $\sigma$ is an element of the symmetric group $S_n$ on $n$ elements.  For the multigluon state it is of course crucial to apply the same permutation $\sigma$ on both $p$ and $a$.

To compute scattering amplitudes, we work in an ``all-in'' formalism in which all states are incoming, though some may have negative energy.  In this prescription, the $n$ graviton and $n$ gluon amplitudes are
\eq{
\langle 0 | T_{\rm gravity} | p\rangle &= M(p) \, \delta^D \! \left({\textstyle \sum} p \right) \\  \langle 0 | T_{\rm YM} | p , a\rangle &= A(p,a) \, \delta^D\! \left({\textstyle \sum} p\right) ,
}{}
where $T$ and $|0\rangle$ denote the scattering matrix and vacuum state in each theory.

Next, we attempt to formalize \Eq{eq:double_copy_eq} into an explicit equation mapping graviton states to entangled gluon states in a double copy Hilbert space, $\Hil_{\rm YM}\otimes \Hil_{\overline{\rm YM}}$.  Crucially, we choose both gauge theory factors to have the {\it same}  gauge group ${\rm SU}(N)$.  
Now consider a general entanglement ansatz that maps a set of $n$ gravitons from $\Hil_{\rm gravity}$ with momenta $p$ to an entangled superposition of $n$ gluons from $\Hil_{{\rm YM}}$ and $n$ gluons from  $\Hil_{\overline{\rm YM}}$, all with corresponding momenta $p$, 
\begin{eqnarray}
& |p_1\rangle   | p_2 \rangle \! \cdots\!  | p_n\rangle &\\
& \updownarrow & \nonumber \\ 
& \sum\limits_{a_1}\sum\limits_{a_2} \cdots \sum\limits_{a_n}\sum\limits_{\sigma \in S_n}
 |p_1, a_1\rangle   | p_2,a_2 \rangle \! \cdots\!  | p_n, a_n\rangle \otimes |p_1,\! \sigma(a_1)\rangle   | p_2,\! \sigma(a_2) \rangle  \!\cdots \! | p_n,\! \sigma(a_n)\rangle K\!(\sigma, \!p_1, \!p_2,\!\ldots\!,\! p_n)  .& \nonumber
\end{eqnarray}
As the proliferation of labels and summations will soon become unwieldy, let us utilize the shorthand in Eqs.~\eqref{eq:grav_ket} and \eqref{eq:gauge_ket} to recast this in a more abbreviated form,
\eq{
| p \rangle \leftrightarrow \sum_{a,\sigma} |p,a\rangle \otimes |p,\sigma(a)\rangle K(\sigma, p).
}{eq:ent_ant}
As noted earlier, the entanglement ansatz in \Eq{eq:ent_ant} implicitly assumes that color indices are only contracted {\it across} copies, i.e., that no YM color indices are contracted with each other and similarly for $\overline{\rm YM}$. We make this choice to draw a parallel with the structure of Kaluza-Klein theories, where the graviphoton states that are ultimately identified as gluons carry internal color indices that link up with the indices of the Killing vector generators on an internal manifold.  In the present context, this simplifying assumption implies that the entangled state ``spontaneously breaks'' the double copy gauge group down to the diagonal ${\rm SU}(N)$.

It bears emphasizing that \Eq{eq:ent_ant} is not a physical quantum field theoretic state in the usual sense due to a several notable peculiarities.  First, we have augmented the Hilbert space to include the negative-energy modes required in the all-in formulation. It would be interesting to see whether one could construct a mapping that relates gravitons to entangled gluons of strictly positive energy, but this will require taking a subset of the particles to be out-states. We leave such a possibility to future analysis.  Second, \Eq{eq:ent_ant} is not a properly unit-normalized wave function.  Naively, this is remedied  by just dividing by the appropriate kinematic-dependent normalization factors, but as we will see later on this is not so simple.
We also note that the state map in \Eq{eq:ent_ant} does not have a tensor product structure: the mapping between graviton and gluon states for $n+m$ particles is {\it not} given by the product of the mapping for $n$ and $m$ particles.

The entanglement kernel $K$ is, a priori, a general function of the permutation $\sigma$ and the external momenta $p$.  However, it can be constrained since we require Bose symmetry on the gravitons.  Thus, we can write the right-hand side of \Eq{eq:ent_ant} as 
\eq{
| p \rangle= | \tau(p) \rangle \leftrightarrow \sum_{a,\sigma} |\tau(p),a\rangle \! \otimes \! |\tau(p),\sigma(a)\rangle K(\sigma, \tau(p)) = \sum_{a,\sigma} |p,a\rangle \! \otimes\! |p,\sigma(a)\rangle K(\tau \sigma \tau^{-1},\tau (p))
}{}
for any $\tau \in S_n$, where in the second equality we have relabeled the sum over $a$ and $\sigma$.
Since each state $ |p, \sigma (a)\rangle $ in the final line is independent, we learn that the entanglement kernel satisfies
\eq{
K(\sigma, p) = K(\tau \sigma \tau^{-1}, \tau(p))
}{eq:perm_kernel}
for any $\tau\in S_n$.  
A trivial corollary to \Eq{eq:perm_kernel} is that
\eq{
K(\sigma, p) = K( \sigma , \tau(p)) \quad \forall \;\, \tau \;\, \textrm{such that} \;\, [\sigma,\tau]=0.
}{eq:K_id}
A special case of this condition is when $\tau=\sigma$, for which it follows that the kernel is invariant under its own corresponding permutation.

Another byproduct of \Eq{eq:perm_kernel} is that any entanglement kernel $K$ is naturally classified according to the conjugacy classes of the permutation group.  For this reason it suffices to fix $K$ for a single representative of a given conjugacy class.  Then by applying \Eq{eq:perm_kernel}, we can obtain $K$ for all other members of that conjugacy class.

\section{Scattering Amplitudes}\label{sec:amplitudes}

We are now equipped to compute the scattering amplitude corresponding to the entangled gluon state defined in \Eq{eq:ent_ant}.   We apply the double copy map to the scattering matrix,
\eq{
T_{\rm gravity} &\leftrightarrow T_{{\rm YM}\otimes \overline{\rm YM}},
}{}
which together with \Eq{eq:ent_ant} implies that
\eq{ 
\langle 0|  T_{\rm gravity}| p\rangle  \leftrightarrow
 \langle 0| \left( T_{{\rm YM}\otimes \overline {\rm YM}} \, \sum_{a,\sigma} |p,a\rangle \otimes |p,\sigma(a)\rangle K(\sigma, p) \right).
}{eq:scattering_mat}
We have thus mapped the $n$-graviton scattering amplitude to a sum over products of $n$-gluon scattering amplitudes,
\eq{
M(p) = {\rm Vol}_{D} \sum_{a,\sigma} A(p,a) \bar{A}(p,\sigma(a)) K(\sigma, p),
}{eq:gauge_to_grav}
where $A$ and $\bar A$ are amplitudes in YM and $\overline{\rm YM}$.  As usual, we can think of $A$ and $\bar A$ as functions of distinct polarization vectors, $\epsilon_\mu$ and $\bar \epsilon_{\bar \mu}$, respectively, whose outer product is the polarization tensor $\epsilon_\mu \bar \epsilon_{\bar \mu}$ of the multiplet describing the graviton, dilaton, and two-form.
The volume of spacetime, ${\rm Vol}_D = \delta^D(0)$, appears because each gauge theory copy produces its own separate momentum-conserving delta function.  In order for both sides of \Eq{eq:gauge_to_grav} to have the same mass dimension, the entanglement kernel must have momentum dependence of the form
\eq{
K \propto \frac{1}{{\rm Vol}_D} \left( \kappa\over g \bar g \right)^{n-2} p^{2(n-3)},
}{eq:p_scaling}
where $\kappa=\sqrt{8\pi G}$, $g$, and $\bar g$ are the gravitational and gauge couplings for the ${\rm YM}$ and $\overline{\rm YM}$ gauge theories in general dimension.  As is standard in analyses of the double copy we will henceforth drop all coupling constants, as well as the ${\rm Vol}_D$ factor.  These factors all come along for the ride and are trivial to reintroduce where needed.

As noted previously, the wave function defined in \Eq{eq:ent_ant} is not properly normalized.  Crucially, dividing the inner product in \Eq{eq:scattering_mat} by the appropriate kinematic normalization factor will necessarily modify the momentum dependence of the resulting amplitude.  In fact, on general grounds, any properly normalized superposition of gluons will be weighted by dimensionless coefficients, which by dimensional analysis cannot grow in the high-energy limit.  Any such normalized state will never mimic the well-known high-energy growth of graviton scattering amplitudes.  For this reason, we do not divide by a normalization factor  in \Eq{eq:scattering_mat}, and \Eq{eq:ent_ant} should instead be interpreted as one non-normalized branch of the wave function.  The remaining support of the wave function must reside in some other branch, which we can choose to be an auxiliary state on $\Hil_{\rm YM}\otimes \Hil_{\overline{\rm YM}}$ that is orthogonal to the state defined in \Eq{eq:ent_ant}.  For example, this other branch could be the vacuum state or perhaps some other exotic mode.  In any case, the S-matrix implicitly contains a projector that zeroes out all but the desired scattering state.   Since we work in an effective theory of gravity where all momenta are below the Planck scale, the normalization of the entangled state defined in \Eq{eq:ent_ant} will always be less than one.

As we will see, it will often be convenient to write the color-dressed amplitude $A(p,a)$ in terms of color-ordered amplitudes, 
\eq{
A(12\cdots n) = A(p_1, p_2, \ldots, p_n),
}{} 
and the half ladder products of the DDM decomposition \cite{DDM},
\eq{
F(a) =  (F^{a_2} F^{a_3} \cdots F^{a_{n-1}})^{a_1 a_n},
}{}
where $(F^{a_2})^{a_1a_3} = i f^{a_1 a_2 a_3}$ is the ${\rm SU}(N)$ generator in the adjoint representation. The relation between color-dressed and color-ordered amplitudes is then: 
\eq{
A(p,a) = \sum_{G\in S_{n-2}} A(G)F(G(a)),
}{eq:DDMform}
where $G$ is a permutation on the set $(1,2,\ldots, n)$ that leaves 1 and $n$ fixed.  Plugging the DDM form of the amplitude back into \Eq{eq:gauge_to_grav}, we obtain:
\eq{
M(p) = {\rm Vol}_{D} \sum_{L,R \in S_{n-2}}  A(L)\bar{A}(R) \sum_{a,\sigma} F(a)F(\sigma(a)) K( R^{-1} \sigma L,p).
}{eq:M_gen}
It is a mechanical exercise to compute the various products of half-ladders $\sum_a F(a) F(\sigma(a))$ for all $\sigma$ in ${\rm SU}(N)$ theory using the usual diagrammatic algorithms \cite{Lance}.  Once the kernel $K$ is specified, we then insert the known expressions for the color-ordered gluon amplitudes to obtain the final scattering amplitude corresponding to the entangled state defined in \Eq{eq:M_gen}.  Finally, we compare this quantity directly to the known graviton scattering amplitude computed via standard methods.

\section{Three-Particle States} \label{sec:3pt}

As we noted previously, it suffices to define the entanglement kernel for a single representative from each conjugacy class of the permutation group.  The kernels for all other elements of that conjugacy class are then obtained through \Eq{eq:perm_kernel}.  For three-particle scattering, we choose the representatives that in permutation cycle notation are $(1)(2)(3)$, $(12)(3)$, and $(123)$. According to the momentum scaling in \Eq{eq:p_scaling}, the kernels are simply momentum-independent constants,
\eq{
K_{(1)(2)(3)} &=\alpha_1 \\
K_{(12)(3)} &=\alpha_2 \\\
K_{(123)} &= \alpha_3.
}{}
Plugging into \Eq{eq:gauge_to_grav}, we obtain the color-dressed form of the amplitude, 
\be 
\begin{aligned}
M(p_1,p_2,p_3) &= A(1^a 2^b 3^c) \left[
\frac{\alpha_1}{6}  \bar A(1^a 2^b 3^c)\! + \! \frac{\alpha_2}{2} \bar A(1^a 3^b 2^c) \! + \! \frac{\alpha_3}{3}  \bar A(3^a 1^b 2^c ) \right]  \!+ \textrm{perm},
\end{aligned}\label{eq:3ptcolor}
\ee
where the right-hand side includes a sum over all permutations on the external momenta and polarizations for legs 1, 2, and 3 but with color indices held fixed.   To obtain this form, we used the identity in \Eq{eq:Bose} that follows from Bose symmetry of the gauge theory amplitudes.   
Next, we simplify the expressions by rewriting the color-dressed amplitudes in terms of  color-ordered amplitudes in \Eq{eq:DDMform}, yielding the general formula in \Eq{eq:M_gen}.  For the case of three-particle scattering, this formula becomes
\eq{
M(p_1,p_2,p_3) =  - N(N^2-1)  \left( \alpha_1- 3 \alpha_2 +2  \alpha_3 \right)  A(123)\bar{A}(123) .
}{eq:M_3}  
As is well known, the product of three-particle amplitudes from YM and $\overline{\rm YM}$ is proportional to the three-graviton amplitude. Unsurprisingly, any generic entangled superposition of gluons successfully reproduces gravity for three-particle scattering.

Note that \Eq{eq:3ptcolor} can be interpreted as a color-dressed, permutation-invariant form of the three-point KLT relations.
For the choice $\alpha_2 = \alpha_3 = 0$, we can write these color-dressed KLT relations in the particularly simple form $M(p_1,p_2,p_3)= A(1^a 2^b 3^c)\bar A(1^a 2^b 3^c)$.

\section{Four-Particle States} \label{sec:4pt}

For four-particle scattering there are five conjugacy classes, and their representative elements are $(1)(2)(3)(4)$, $(12)(3)(4)$, $(12)(34)$, $(123)(4)$, and $(1234)$. By \Eq{eq:p_scaling}, each kernel must be a linear function of $s$, $t$, and $u$ that furthermore must satisfy \Eq{eq:K_id}.\footnote{Throughout, we work in mostly-minus metric signature where $s=(p_1+p_2)^2$, $t=(p_1+p_4)^2$, $u=(p_1+p_3)^2$, and more generally $s_{ij} = (p_i+p_j)^2$.}  These constraints immediately imply that
\eq{
K_{(1)(2)(3)(4)} &= 0  \\
K_{(123)(4)} &= 0 .
}{}
The first kernel is a linear function of $s$ and $t$ that is invariant under all permutations, and the only such combination is $s+t+u=0$.  Similarly, the second kernel is a linear function of $s$ and $t$ that is invariant under the cyclic permutation $(123)$, again yielding zero.

Applying similar logic, we find that the nontrivial kernels are
\eq{
K_{(12)(3)(4)} &= \beta_1  s \\
K_{(12)(34)} & = \beta_2 s \\
K_{(1234)} & = \beta_3 u,
}{eq:K_4}
where the first and second lines are the only possible functions invariant under the transposition $(12)$, the third line is the only possible function invariant under cyclic permutation $(1234)$, and the $\beta_i$ are momentum-independent constants.  

Again evaluating \Eq{eq:gauge_to_grav}, we obtain the four-particle amplitude for this entangled state expressed in terms of color-dressed amplitudes,
\be
\begin{aligned}
M(p_1,p_2,p_3,p_4) =   s A(1^a 2^b 3^c 4^d) \big[ &\phantom{{}+{}}\beta_1  \bar A(1^a 2^b 4^c 3^d ) + \beta_1 \bar A(2^a 1^b 3^c 4^d) + \beta_2  \bar A(2^a 1^b 4^c 3^d ) \\
&  + \beta_3  \bar A(3^a 4^b 2^c 1^d) + \beta_3 \bar A(4^a 3^b 1^c 2^d )  \; \big] \\& \hspace*{-2.62cm} +  \text{$t$-channel}+ \text{$u$-channel}.
\end{aligned}\label{eq:4ptcolor}
\ee
Here the $t$-channel and $u$-channel terms are obtained by permuting the momenta and polarizations of the $s$-channel term, written out explicitly above, by $1\leftrightarrow 3$ and $1\leftrightarrow 4$, respectively.  The reason for this nomenclature is that the $s$-channel term manifestly exhibites the symmetries of an $s$-channel cubic diagram, and similarly for the other channels.
As before, we can again apply \Eq{eq:M_gen} to obtain the same expression in terms of color-ordered amplitudes, 
\eq{
M(p_1,p_2,p_3,p_4) = -3 N^2 (N^2-1)  \left(\beta_1 + \beta_3 \right)  
\big[&\phantom{{}-{}} s A(1234)\bar{A}(1234) -t A(1324)\bar{A}(1234) \\
&-t A(1234)\bar{A}(1324) +u A(1324)\bar{A}(1324)\;\;\big],
}{eq:M_4}
so $\beta_2$ actually drops out from the physical answer.
We can verify explicitly that \Eq{eq:M_4} is exactly proportional to the four-particle scattering amplitude in gravity.   In particular, plugging in the BCJ relations,
\eq{
A(1324) &= \frac{s}{u} A(1234) \\
 \bar{A}(1234) &= \frac{u}{s} \bar{A}(1324),
 }{}
we find that the right-hand side of \Eq{eq:M_4} is proportional to $t A(1234) \bar{A}(1324)$, which is precisely the four-particle KLT relation.  As a corollary, we can interpret \Eq{eq:4ptcolor} as a color-dressed,  permutation-invariant form of the four-particle KLT relations.  That the resulting object is even a sensible  amplitude is quite surprising since the product of YM and $\overline{\rm YM}$ amplitudes has numerous spurious poles that must simultaneously cancel.

We thus arrive at the remarkable conclusion that any arbitrary entangled superposition of gluons---subject to our minimal assumptions discussed previously---necessarily scatters like gravitons.  Note that this is highly nontrivial: our entanglement ansatz guarantees gauge invariance and permutation invariance of the resulting amplitude, but {\it not} locality.  Here we see that the latter is a byproduct of the other conditions.

From \Eq{eq:M_4}, it is clear that one can construct an elegantly simple entangled gluon state that reproduces gravity, for any choice of entanglement ansatz coeffcients provided $\beta_1 + \beta_3 \neq 0$ so the resulting amplitude does not vanish.  For example, one can choose $\beta_2 = \beta_3 = 0$, leaving only a transposition, or choose $\beta_1 = \beta_2 = 0$, leaving only a cyclic permutation.

An even more bottom-up understanding is obtained by building a completely arbitrary quadratic form of the ${\rm YM}\otimes \overline{\rm YM}$ amplitudes that appear in \Eq{eq:M_4}, with coefficients given by linear polynomials in $s$, $t$, and $u$. It is then straightforward to verify that by imposing permutation invariance on the external kinematic variables, the resulting object is necessarily proportional to the known graviton amplitude.

\section{Five-Particle States} \label{sec:5pt}

In the five-particle case there are seven conjugacy classes.  As before, we select a representative permutation from each class in order to define the kernels and then apply \Eq{eq:K_id} to obtain all others.  For example, since $K_{(1)(2)(3)(4)(5)}$ corresponds to the identity permutation, its kinematic dependence must be invariant under all permutations.  After accounting for on-shell conditions, the only such function at ${\cal O}(p^4)$ is $s_{ij}^2$ summed over all $i$ and $j$.    Applying the same logic to the other kernels, we arrive at a total of twenty-two free coefficients in the entanglement ansatz:  
\be 
\begin{aligned}
K_{(1)(2)(3)(4)(5)} &=\phantom{+} \gamma_{1}\sum_{i=1}^{5}\sum_{j=1}^{5}s_{ij}^{2} \\
K_{(12)(3)(4)(5)} & =\phantom{+}\gamma_{2}\left[(s_{13}+s_{14})^{2}+(s_{23}+s_{24})^{2}-s_{13}s_{14}-s_{23}s_{24}-s_{12}(s_{12}+s_{34})\right]\\
 & \phantom{=}+\gamma_{3}\left[s_{13}(2s_{23}+s_{24})+s_{14}(s_{23}+2s_{24})-s_{12}s_{34}\right]\\
 &  \phantom{=}+\gamma_{4}s_{12}^{2}\\
K_{(123)(4)(5)} & =\phantom{+} \gamma_{5}\left[s_{14}(2s_{15}+s_{25})+s_{24}(s_{15}+2s_{25})-s_{12}s_{45}\right]\\
 &  \phantom{=} +\gamma_{6}(s_{12}s_{13}+s_{12}s_{23} + s_{13} s_{23})\\
  &  \phantom{=} +\gamma_{7}s_{45}^{2}\\
K_{(12)(34)(5)} & = \phantom{+} \gamma_{8}\left[(s_{12}+s_{23})(s_{23}+s_{34})-s_{14}s_{24}-s_{13}(s_{14}+s_{24})\right]\\
 &  \phantom{=} +\gamma_{9}(s_{13}s_{24}+s_{14}s_{23}-s_{12}s_{34})\\
&  \phantom{=} +  \gamma_{10}(s_{12}^{2}+s_{34}^{2})\\
 &  \phantom{=} +\gamma_{11}s_{12}s_{34}\\
K_{(123)(45)} & =\phantom{+} \gamma_{12}\left[(s_{12}\! +\! s_{14})^{2}\! + \! (s_{12}\! +\! s_{24})^{2}\! -\! s_{12}^{2}\! +\! s_{14}(s_{13}\! - \! s_{34})\! +\! s_{24}(s_{23} \! - \! s_{34})\! - \! s_{13}s_{23}\right]\\
 &  \phantom{=} +\gamma_{13}\left[s_{14}(s_{15}+s_{24})+(s_{13}+s_{34})(s_{24}-s_{34})-s_{23}s_{34}\right]\\
 & \phantom{=} +\gamma_{14}s_{45}^{2}\\
K_{(1234)(5)} &=\phantom{+} \gamma_{15}\left[(s_{12}+s_{13})^{2}+(s_{13}+s_{34})^{2}-s_{13}^{2}-s_{14}s_{23}+s_{12}s_{34}\right]\\
 &  \phantom{=} +\gamma_{16}(s_{14}+s_{23}+s_{24})(s_{13}-s_{24})\\
 &  \phantom{=} +\gamma_{17}(s_{13}-s_{24})^{2}\\
 &  \phantom{=} +\gamma_{18}(s_{12}s_{34}+s_{14}s_{23})\\
 &  \phantom{=} +\gamma_{19}s_{13}s_{24}\\
K_{(12345)} & =\phantom{+} \gamma_{20}\left[s_{12}(s_{12}+s_{15}-s_{35})+s_{15}(s_{15}+s_{25})-s_{23}(s_{25}+s_{35})\right]\\
 &   \phantom{=} +\gamma_{21}\left[(s_{15}+s_{25})(2s_{15}-s_{23}+s_{25}+s_{35})+s_{15}s_{35}+s_{12}(-2s_{23}+s_{25}+s_{35})\right]\\
 &  \phantom{=} +\gamma_{22}\left[(s_{13}+s_{23})^{2}+(s_{23}+s_{25})^{2}-s_{12}(s_{12}+2s_{15})+2s_{23}s_{35}\right].
\end{aligned}\label{eq:5ptkernels}
\ee
Plugging the kernels in \Eq{eq:5ptkernels} into \Eq{eq:M_gen}, we obtain a lengthy expression that can be checked against the known graviton scattering amplitude.  In contrast with three- and four-particle scattering, here we discover that the entanglement ansatz does {\it not} automatically reproduce the known gravity amplitude.  
Instead, a match occurs for a subspace of the original entanglement ansatz defined by a set of homogenous constraints, $\vec v_{1,2,3,4} \cdot \vec \gamma = 0$, where
\be 
\setlength{\tabcolsep}{2pt}
\begin{tabular}{r r r r r r r r r r r r r r r r r r r r r r r}
$\vec v_1=$ & $(8,$& $0,$& $0,$& $0,$ & $6,$ & $0,$ & $3,$ &  $1,$ & $0,$ & $-7,$ & $-1,$ &  $0,$ & $0,$ & $0,$ & $0,$ & $0,$ &  $0,$ &  $0,$ &  $0,$ & $5,$ & $4,$ & $1)$\phantom{,}\\
$\vec v_2=$ &$(0,$ & $4,$ & $0,$ & $-4,$ & $14,$ & $0,$ & $7,$ & $7,$ &  $6,$ & $-7,$ & $-3,$ & $-4,$ & $8,$ & $0,$ & $-4,$ & $0,$ & $0,$ & $4,$ & $4,$ & $1,$ & $12,$ &  $5)$\phantom{,}\\
$\vec v_3=$ &$(0,$ & $0,$ & $8,$ & $-4,$ & $10,$ & $0,$ & $5,$ & $5,$ & $10,$ & $-5,$ & $-5,$ & $8,$ & $0,$ & $-8,$ & $4,$ & $4,$ & $-8,$ & $12,$ & $4,$ & $-5,$ & $20,$ & $15)$\phantom{,} \\
$\vec v_4=$ &$(0,$ & $0,$ & $0,$ & $0,$ & $0,$ & $0,$ & $0,$ & $0,$ & $0,$ & $0,$ & $0,$ & $0,$ & $0,$ & $0,$ & $1,$ & $-1,$ & $0,$ & $0,$ & $0,$ & $0,$ & $0,$ & $0)$,
\end{tabular}
\label{eq:homo}
\ee
together with a single inhomogeneous constraint $\vec v_0 \cdot \vec \gamma \neq 0$, where
\be 
\setlength{\tabcolsep}{2pt}
\begin{tabular}{r r r r r r r r r r r r r r r r r r r r r r r r r}
$\vec v_0 = $ & $(0,$ & $0,$ & $2,$ &  $-1,$ & $0,$ & $0,$ & $0,$ & $0,$ & $0,$ & $0,$ & $0,$ & $2,$ & $0,$ & $-2,$ & $1,$ & $1,$ & $-2,$ & $3,$ & $1,$ & $0,$ & $0,$ & $0)$,
\end{tabular}\label{eq:inhomo}
\ee
writing $\vec \gamma$ as the vector whose components are $\gamma_i$.  
The four homogenous constraints in \Eq{eq:homo} correspond to contributions to the entanglement ansatz that contribute nontrivially to the final scattering amplitude but in a way that is inconsistent with the correct answer from gravity.  Consequently, these directions must be zeroed out.  Within the remaining eighteen-dimensional subspace of matching solutions, seventeen altogether evaporate from the final amplitude, so their corresponding coefficients are unfixed.  Finally, at least one linear combination, denoted by the final inhomogenous constraint in \Eq{eq:inhomo}, must be present with nonzero coefficient so that the resulting amplitude agrees with gravity and is nonzero.

It is well motivated to ask whether there exist any natural or simple conditions on the twenty-two-parameter, permutation-invariant entanglement ansatz that will automatically reproduce gravity. As it turns out, there are several options of this kind that involve well-known selection rules on amplitudes involving gravitons, dilatons, and two-forms.  For example, for tree-level scattering processes without external two-forms, we can drop these fields from the discussion since they never appear internally due to spacetime parity.  Focusing on the graviton and dilaton sector, we then go to Einstein frame and apply a dilaton field redefinition to rewrite the action as that of a dilaton coupled minimally to Einstein gravity. This theory exhibits a manifest dilaton parity that implies that all amplitudes with an odd number of dilatons, no two-form fields, and any number of gravitons will vanish.  Applying this condition to the twenty-two-parameter entanglement ansatz, one obtains the first three homogeneous constraints in \Eq{eq:homo}, 
which are actually sufficient for the resulting amplitudes to agree, modulo normalization, with the known ones for gravitons and dilatons. 

Another option is to utilize the underlying ${\rm U}(1)$ duality of gravitational amplitudes in four spacetime dimensions. In this case the two-form field may be dualized to an axion, which together with the dilaton forms a complex scalar field $\phi$.  From the viewpoint of the double copy, $\phi$ is the product of a negative helicity gluon from YM and a positive helicity gluon from $\overline {\rm YM}$.  The ${\rm U}(1)$ duality symmetry then dictates that all amplitudes with unequal numbers of $\phi$ and $\phi^*$ particles should vanish.  Imposing this constraint on five-particle scattering is equivalent to demanding that the double copy of the MHV and $\overline{\rm MHV}$ amplitude is zero.  Remarkably, fixing this ${\rm U}(1)$ duality constraint is sufficient to reduce the full twenty-two-parameter entanglement ansatz down to the correct eighteen-dimensional space defined by Eqs.~\eqref{eq:homo} and \eqref{eq:inhomo}, which accords with all amplitudes of gravitons, dilatons, and two-form fields.

All together, this result offers evidence that permutation invariance and gauge invariance---together with certain parity and duality constraints---are sufficient conditions to fix gravity {\it without} assuming locality.  This is reminiscent of an earlier result \cite{Nima+Rodina} proving that tree-level factorization follows from gauge invariance and locality while also conjecturing that locality follows from gauge invariance.  Obviously, the latter conjecture hinges on a number of important caveats---otherwise one could take the KLT relations, which automatically enforce gauge invariance, and arbitrarily modify the momentum kernels and still obtain gravity.
In particular, the conjecture of \Ref{Nima+Rodina} does not overlap or apply to the case studied here, since the former assumes $n-3$ propagator denominators in the amplitude, while products of amplitudes generically give twice as many.

Last but not least, we observe that there exists a choice of ansatz parameters that defines a particularly elegant and simple entangled gluon state that reproduces the five-graviton amplitude.  This occurs for $\gamma_8 = \gamma_9 = \gamma_{11} = 2\gamma_4/5$, with all other $\gamma_i$ vanishing, in which case the color-dressed five-point relations are:
\begin{eqnarray}
& M(p_1,p_2,p_3,p_4,p_5)& \label{eq:simple5} \\
&  \verteq & \nonumber \\
& \gamma_4 A(1^a 2^b 3^c 4^d 5^e) \left[ \frac{1}{12}s_{12}^2 \bar A(2^a 1^b 3^c 4^d 5^e)  - \frac{1}{20} (s_{12}\! + \! s_{13} \! + \! s_{24})(s_{12} \! + \! s_{14} \! + \! s_{23})\bar A(2^a 1^b 4^c 3^d 5^e)  \right] + \text{perm}.& \nonumber 
\end{eqnarray}
Similar to \Eq{eq:3ptcolor},  this expression is summed over all permutations on the external momenta and polarizations for legs 1 through 5. Note that the $(12)(3)(4)(5)$ and $(12)(34)(5)$ conjugacy classes are the only nontrivial contributions to \Eq{eq:simple5}.
It would be very interesting to see if similarly simple expressions can also be found for six-particle scattering and higher. We leave this possibility for future study.

\section{Entanglement Entropy} \label{sec:entanglement}

Equipped with an explicit mapping between gravitons and entangled gluons, it is natural to ask about the entropic properties of the corresponding state.  To this end, we compute the density matrix for the gluon states defined in \Eq{eq:ent_ant}.  Tracing over the $\overline {\rm YM}$ gauge theory, we obtain the reduced density matrix,
\eq{
\rho &= \overline{\rm Tr} \left(  \sum_{a,b,\sigma,\tau} K(\sigma, p) K^*(\tau, p) \;  |p,a\rangle  \langle p,b| \otimes |p,\sigma(a)\rangle    \langle p,\tau(b)|   \right) \\
& = \sum_{a,\sigma, \tau} K(\sigma, p) K^*(\tau, p) |p,\sigma(a) \rangle \langle p,\tau(a)| .
}{eq:rho}
As noted previously, $\rho$ is not properly normalized. Interpreting the entangled state in \Eq{eq:ent_ant} as but one branch of a larger wave function, any expectation value within the YM theory evaluated on this branch should be divided by appropriate factors of ${\rm Tr}(\rho)$, where ${\rm Tr}$ and $\overline {\rm Tr}$ are traces over the YM and $\overline{\rm YM}$ degrees of freedom, respectively.

Ideally, we would now compute the von Neumann entropy of the entangled state, but this would require analytically diagonalizing $\rho$, which is at best extremely difficult and at worst impossible.  A proxy to the von Neumann entropy is the ``purity'' of the state, defined by
\eq{
P &= \frac{ {\rm Tr}(\rho^2) }{ {\rm Tr} (\rho)^2},
}{}
which is closely related to the second-order R\'enyi entropy $S_2(\rho) = - \log P$. On general grounds, $1/d \leq P \leq 1$, where $d$ is the Hilbert space dimension.  The upper and lower bounds on $P$ correspond to pure and maximally-mixed states, respectively.
The linear entropy $1-P$ is a lower bound on the von~Neumann entropy, and can be viewed as the first term in a Taylor expansion about a pure state.

As an additional simplification, in the large-$N$ limit the trace simplifies to
\eq{
{\rm Tr} \left( \sum_a |p,\sigma(a) \rangle \langle p,\tau(a)| \right) =   N^{2n} \delta_{\sigma \tau}  + \cdots,
}{eq:traceN}
where $n$ is the number of particles, the ellipses denote contributions subleading in large $N$, and the Kronecker delta function fixes $\sigma$ and $\tau$ to be the same permutation.   This expression implies that  
\be 
{\rm Tr} (\rho) = N^{2n} \sum_\sigma \left|K(\sigma, p)\right|^2 +\cdots
\ee
and furthermore that
\be 
\begin{aligned}
{\rm Tr} (\rho^2) &={
\rm Tr} \left(\sum_{\substack{a,\sigma,\tau\\a',\sigma',\tau'}} K(\sigma,p) K^*(\tau,p) K(\sigma',p) K^*(\tau',p) \ket{p,\sigma(a)}\langle p,\tau(a)|p,a'\rangle \bra{p,\tau'\sigma'^{-1}(a')}\right)
\\& = N^{2n} \sum_{\sigma,\tau,\sigma'} K(\sigma,p) K^*(\tau,p) K(\sigma',p) K^*(\sigma \tau^{-1} \sigma',p) + \cdots
\\&=N^{2n} \sum_\tau \left| \sum_\sigma K(\sigma, p)K^*(\tau \sigma, p) \right|^2 +\cdots.
\end{aligned}\label{eq:trrho2}
\ee
The first equality in \Eq{eq:trrho2} comes from the definition of $\rho$ in \Eq{eq:rho} after relabeling one of the dummy color indices via multiplication by $\sigma^{-1}$. In the second equality, we set $a'=\tau(a)$ using the Kronecker delta functions from $\langle p,\tau(a)|p,a'\rangle$ and applied \Eq{eq:traceN} to evaluate the trace in the large-$N$ limit. To obtain the third equality, we relabeled the sums over permutations according to $\tau \rightarrow \tau \sigma$ and $\sigma'\rightarrow \tau \sigma'$ and then rewrote the sum over $\sigma'$ as a perfect square. As expected, both ${\rm Tr}(\rho)$ and ${\rm Tr}(\rho^2)$ are manifestly nonnegative.  It is straightforward to derive formulas for higher-order R\'enyi entropies that are similar to the above equations.

We thus find that the purity scales at large $N$ as
\eq{
P \propto N^{-2n},
}{}
where the proportionality factor is dependent on kinematics.  At large $N$, the dimension of the Hilbert space scales as $d \simeq N^{2n}$ since there are $n$ gluons, each with $N^2-1$ color degrees of freedom.  Since the  large-$N$ purity approaches the allowed lower limit, $1/d$, the state is maximally mixed.  In turn, the von~Neumann entropy is $\log d$ up to subleading corrections at large $N$.  We emphasize that the phenomenon of maximal mixing should be expected since the graviton is effectively a composite state built from a large number of gluons.  We also note that factors of spacetime volume and coupling constants in \Eq{eq:p_scaling} all exactly cancel from the purity, irrespective of the large-$N$ limit.

For the case of three-particle scattering, ${\rm Tr}(\rho) \propto {\rm Tr}(\rho^2) \propto N^6$ are independent of the external kinematics and depend only on the coefficients of the entanglement ansatz. 
Notably, in this case one can choose the coefficients $\alpha_i$, for example with $\alpha_2 = \alpha_3 = 0$, to exactly saturate the absolute lower limit for the purity at $1/d$, resulting in a maximally-entangled state at finite $N$. 
For this choice, the three-particle wave function is a manifestly maximally-entangled Bell state of a two-qudit system, where each qudit has Hilbert space dimension $(N^2-1)^3$,
\be
\ket{abc}\ket{abc} = \ket{111}\ket{111} + \ket{112}\ket{112} + \cdots = \sum_i \ket{i}\ket{i},
\ee
where $i$ runs over all configurations of $abc$.  

 Meanwhile, for the four-particle case, we find
\eq{
{\rm Tr}(\rho) &\propto N^8(s^2 + t^2 + u^2) \\
{\rm Tr}(\rho^2) &\propto N^8 (s^2 + t^2 + u^2)^2.
}{eq:rho_4}
The kinematic structure shown above not so surprising given the scarcity of permutation-invariant functions of external momenta permitted in four-particle scattering.  Altogether, this implies that the purity $P$ for three- and four-particle scattering is completely independent of the kinematics.  

On the other hand, by explicit calculation one finds that for five-particle scattering, $P$ is always dependent on kinematics.  Hence $P$ will be extremized for entanglement ansatz parameters in a way that depends on the kinematics chosen.  This immediately eliminates the naive possibility that one might use extremization of $P$ to fix the entanglement ansatz to universal values, perhaps fixing the free coefficients to match gravity automatically.  

\section{Future Directions} \label{sec:conclusions}

The results of this work leave a number of compelling directions for future research.  First and foremost is the question of how our results might be generalized beyond five-particle scattering to arbitrary numbers of external legs.  Obviously, armed with a complete and explicit state-to-state map one would be poised to decipher all manner of gravitational phenomena in terms of the double copy construction.  Such a map could also shed light on the classical double copy \cite{Monteiro:2014cda,Monteiro:2015bna,Luna:2016hge} as well as implementations of color-kinematics duality at the action level \cite{Kiermaier, Donal, flavor-kinematics, pions_as_gluons, twofold}.  Another natural direction would be to better understand the entanglement and complexity properties of these states.

A convenient byproduct of an all-$n$ generalization of the state-to-state map would be a fully permutation-invariant, color-dressed version of the KLT relations.   In fact, this suggests a natural path to deriving such a map, which is by reverse-engineering it from more permutation-invariant forms of the existing KLT relations \cite{BjB}.

More speculatively, it would be interesting to see if there exists any connection between the entangled gluon states we have constructed and the established holographic formulation of gravitational states, for example in the thermofield double describing an entangled CFT state dual to an AdS black hole.

\pagebreak
 
\begin{center} 
{\bf Acknowledgments}
\end{center}
\noindent 
We thank Jake Bourjaily, JJ Carrasco, Matt Headrick, and Junyu Liu for useful discussions and comments.  
C.C. is supported by the DOE under grant no.~DE- SC0011632 and by the Walter Burke Institute for Theoretical Physics.
G.N.R. is supported by the Miller Institute for 
Basic Research in Science at the University of California, Berkeley.

\bibliographystyle{utphys-modified}
\bibliography{entangled_gluons}

\end{document}